\documentclass[aps,prd,twocolumn,nofootinbib,superscriptaddress,10pt,floatfix,longbibliography]{revtex4-2}
\usepackage{amsmath,amsthm,amsfonts,amssymb,amscd,bbold,mathrsfs,bm}
\usepackage{graphicx}
\usepackage{float}
\usepackage{dcolumn}
\usepackage{physics}
\usepackage[hidelinks]{hyperref}
\usepackage{xcolor}
\usepackage{CJK}
\newcommand{\rev}[1]{#1}

\begin{document}
\raggedbottom
\begin{CJK*}{UTF8}{}
\title{Axion signal searching using hybrid nuclear-electronic spin systems}
\author{Xiangjun Tan ({\CJKfamily{gbsn}\mbox{谈相君}})}
\email{xiangjun.tan.25@ucl.ac.uk}
\affiliation{Department of Physics and Astronomy, University College London, WC1E 6BT London, United Kingdom}
\author{Zhanning Wang}
\affiliation{Instituto de Ciencia de Materiales de Madrid, Consejo Superior de Investigaciones Cient\'{\i}ficas, Sor Juana In\'es de la Cruz 3, 28049 Madrid, Spain}
\date{September 22, 2026}

\begin{abstract}
Conventional nuclear magnetic resonance searches for the galactic axion wind face technical challenges at lower frequencies, where the sensitivity of inductive readout scales less efficiently.
In the pulsed architecture, a nuclear control sequence filters the axion field, a final $\pi/2$ pulse maps phase onto population, and the electron spin reads the resulting hyperfine-conditioned shift.
We demonstrate analytically that this dispersive upconversion preserves the specific sidereal and annual modulation signatures required to distinguish dark matter signals from instrumental backgrounds.
For a silicon ${ }^{209}$Bi donor platform, the projected hybrid sensitivity exceeds that of direct nuclear detection by more than an order of magnitude over the mass range $10^{-16}$\textendash{}$10^{-10}$~eV.
With additional collective and resonant enhancements, the projected sensitivity can reach discovery-level Dine-Fischler-Srednicki-Zhitnitsky-scale axion-nucleon couplings under the assumed integration time.
\end{abstract}
\maketitle
\end{CJK*}

\textit{Introduction.-}
Searches for ultralight dark matter, including axions and axion-like particles (ALPs), span quantum sensing and particle physics~\cite{Kim2005, Arvanitaki2010, Graham2015, MARSH2016, Irastorza2018, Luca2020, Choi2021, Fujita2021, Ken2021, Adachi2023, Ahn2024, Altenmuller2024, Ghosh2024, Manzari2024, Garcia2025, Goodman2025, Berezhiani:1989fp, Khlopov:1999tm, Bi:2022yyg, l8kd-yhlg}.
For axions/ALPs with derivative couplings to nucleon spins, the field acts as an effective axion wind that drives nuclear spins at the Compton frequency $\omega_a \simeq m_a c^2/\hbar$ and produces sidereal and annual modulations in the laboratory frame~\cite{Graham2013, Sikivie2021}.
These directional modulations provide an astrophysical test ground against terrestrial backgrounds.

Nuclear magnetic resonance (NMR) searches, exemplified by CASPEr-Wind, exploit large nuclear polarization and long coherence to set leading bounds on axion\textendash{}nucleon wind couplings~\cite{Budker2014, Garcon2018, PhysRevLett.131.091002, Stadnik2015_EPJC, universe2022}. 
Axion-driven sample magnetization can be read out inductively, where the pickup voltage scales as $V\propto \mathrm{d}\Phi/\mathrm{d}t\propto \omega_a$, or via precision magnetometry (e.g., SQUIDs or atomic magnetometers) that measures flux or field more directly. In the coherent averaging regime, the optimal coherent integration time is limited by the axion coherence time $\tau_a\propto 1/\omega_a$, so the explicit $\omega_a$ in inductive voltage pickup partly cancels and, for an otherwise white noise floor, $\mathrm{SNR}\propto \omega_a\sqrt{\tau_a}\sim \sqrt{\omega_a}$.

Compared with nuclei, electron spins have a much larger gyromagnetic ratio and benefit from mature, high-sensitivity readout, including spin-to-charge conversion via Pauli spin blockade and gate or microwave-based dispersive reflectometry~\cite{Elzerman2004, Hanson2005, Degen2017, Crippa2019}.
In donor and quantum dot platforms the strong hyperfine interaction $A\bm{I}\cdot\bm{S}$ tightly couples nuclear and electronic degrees of freedom, enabling nuclear dynamics to be mapped onto charge or electron-spin observables with high fidelity~\cite{Morton2008, Muhonen2014, Pla2013, Pla2021}.
Recent spin-echo experiments in Ge hole-spin qubits have resolved hyperfine-mediated nuclear-bath dynamics through the qubit response, directly demonstrating nuclear-to-spin-qubit transduction in a semiconductor platform~\cite{Zeng2026}.
In a hybrid sensor, the nuclear ensemble is the axion-sensitive element and a coupled electron provides fast readout.
Such a sensor must bridge the nuclear and electronic timescales and energy scales while transducing the nuclear response without degrading its spatiotemporal coherence or sidereal and annual signatures.

Here we develop a hyperfine-mediated upconversion readout for axion-wind searches.
As a benchmark, we quote sensitivity relative to the Dine-Fischler-Srednicki-Zhitnitsky (DFSZ) class of QCD axion models, which arise in two-Higgs-doublet extensions of the Standard Model and predict axion couplings to Standard-Model fermions set by the Peccei\textendash{}Quinn scale.
The target is the axion\textendash{}nucleon wind coupling that drives nuclear spins; the electron spin serves as a transduction/readout channel via hyperfine-mediated upconversion rather than as the primary axion-coupled sensor.

Building on Refs.~\cite{Xiangjun2025, Xiangjun2026_QEC, Xiangjun2025_2}, we extend the approach to a nuclear\textendash{}electronic hybrid architecture with collective enhancement.
We construct a filter function formalism for nuclear\textendash{}electronic transduction, defining a hybrid gain $G_{\text{hyb}}$ and nuclear filter function $Y_N(\omega)$ that give platform-independent expressions for sensitivity.
We show that hyperfine upconversion preserves the axion-wind kinematics: after demodulation, the electronic record contains a sidereal component at $\Omega_\star$, a pure annual component at $\Omega_\oplus$, and mixed sidebands at $\Omega_\star\pm\Omega_\oplus$, providing a geometric veto against terrestrial systematics. When the axion coherence time is shorter than the observation baseline, these harmonics describe deterministic modulation of segment amplitudes or covariance rather than a single phase-coherent waveform over the complete run. We evaluate the approach in a silicon ${}^{209}\text{Bi}$ donor system, incorporating realistic coherence times, control pulse constraints, and readout noise to benchmark the projected performance against existing NMR strategies.
We use Si:$^{209}$Bi donors as the benchmark because their large isotropic hyperfine coupling enables strong transduction alongside established control and long coherence in enriched silicon.

The hybrid readout can improve the readout-limited sensitivity relative to direct nuclear detection across the accessible mass range. \rev{With collective nuclear-spin enhancement and a high-$Q$ resonator, the idealized benchmark overlaps part of the DFSZ reference region for a one-year dwell at each scanned mass point.}
At low masses the overall reach of wind searches is often limited by the readout chain rather than by nuclear coherence alone, particularly for pickup-coil \emph{voltage} detection with its explicit $\omega_a$ transduction factor.
Hyperfine-mediated nuclear-to-electron transduction removes this explicit inductive penalty by mapping the axion-driven nuclear response onto a high-bandwidth electronic readout channel, leaving the remaining frequency dependence set by $\tau_a$ and by the control-sequence filter functions.

\textit{Model and theory.-}
Unless otherwise stated, we set $\hbar=c=1$ and express Hamiltonians in angular-frequency units. We consider a semiconductor donor system with a spin-1/2 electron $\bm{S}$ and a nuclear spin $\bm{I}$ (for \textsuperscript{209}Bi, $I=9/2$) in a static magnetic field $\bm{B}=B_z\hat{\text{z}}$~\cite{Wolfowicz2013,Yasukawa2016}.
The laboratory-frame Hamiltonian is
\begin{equation}
H_{\text{total}} = \omega_e S_z + \omega_N I_z + A \bm{S} \cdot \bm{I} + H_{aN}(t).
\end{equation}
The electron and nuclear Larmor frequencies are $\omega_e=|\gamma_e| B_z$ and $\omega_N=\gamma_N B_z$, where $\gamma_e$ is the electron gyromagnetic ratio. The isotropic hyperfine interaction constant $A$ is taken from experiment. For a donor-bound nucleus the secular hyperfine term shifts the nuclear precession frequency to $\omega_N^{\rm eff}=\gamma_N B_z\pm A/2$, depending on the electron spin state; for Si:$^{209}$Bi this places the nuclear drive near $737$~MHz rather than at the bare nuclear Zeeman frequency. All sequences below are defined in the corresponding nuclear rotating frame. Because an electron spin flip shifts $\omega_N^{\rm eff}$ by $\sim A$ and thereby destroys the accumulated nuclear phase, the protocol additionally requires $T_1^{(e)}\gg\tau_N$ throughout the nuclear interrogation window.

\rev{The axion\textendash{}nucleon interaction is written as an effective field
$H_{aN}(t)=-\gamma_N\bm{I}\cdot\bm{B}_a(t)$, with
$\bm{B}_a(t)=\bm{B}_{a,0}\cos(\omega_a t+\phi)$.  We use the
dimensionful derivative coupling $g_{aNN}=C_N/f_a$ and the
dimensionless coupling plotted in Fig.~\ref{fig:axion_5sigma_compare_ent_res},
$g_{an}\equiv m_N g_{aNN}$.  In angular-frequency units,
$\bm{B}_a=g_{aNN}\nabla a/(\hbar\gamma_N)
=g_{an}\nabla a/(m_N\hbar\gamma_N)$.}
The fixed Galactic-frame direction of $\bm{B}_a$ leads to sidereal and annual modulations in the lab frame, and the drive remains coherent for a time $\tau_a$ set by the velocity dispersion.
The coherence time of this drive is characterized by $\tau_a \simeq 2 \pi /\left(m_a v_0^2\right)$ under the standard halo model (SHM).

We focus on the longitudinal component of the axion-induced field, $B_{a,z}(t)$, which modulates the nuclear Larmor frequency.
To cover a broad frequency range, we use standard Ramsey/echo and multipulse sequences (e.g., Ramsey, Hahn echo, XY8, and Carr\textendash{}Purcell\textendash{}Meiboom\textendash{}Gill (CPMG)) to shape the nuclear spectral response~\cite{Medford2012,Wang2012,Farfurnik2015}.
The spectral sensitivity of a sequence is characterized by its filter function $Y_N(\omega)$, the Fourier transform of the time-domain sensitivity function $y_N(t) \in \{0, \pm 1\}$ over the sensing duration. The values $\pm1$ correspond to free-precession segments whose sign is toggled by pulses, while $0$ accounts for intervals with negligible phase accumulation.
For a narrowband axion field $B_{a,z}(t) = B_a \cos(\omega_a t + \phi)$, the accumulated phase $\varphi_N$ at the end of the sequence is given by
\begin{equation}
\varphi_N \approx \gamma_N B_a |Y_N(\omega_a)| \cos(\omega_a t_0 + \Phi_{\text{off}}) \,,
\end{equation}
where $t_0$ marks the start of the sequence and $\Phi_{\text{off}}$ includes the phase of the filter function and the axion field.

The choice of sequence determines the detection window.
The Ramsey sequence ($y_N=1$) exhibits a low-pass filter response $|Y_N(\omega)| \approx \tau \operatorname{sinc}(\omega \tau/2)$, maximizing sensitivity to ultralight axion fields.
Conversely, Hahn echo and CPMG sequences act as band-pass filters, suppressing low-frequency noise while extending sensitivity to higher-frequency (oscillating) axion fields determined by the pulse spacing~\cite{Lutchyn2008, Gonzalo2011, Naydenov2011}.

Experimentally, the nuclear spin is initialized in a superposition state (e.g., along $\hat{\text{x}}$).
After accumulating the axion-induced phase $\varphi_N$ over time $\tau$, a final $\pi/2$ pulse maps this transverse phase onto the longitudinal polarization $\langle I_z \rangle$.
In the small-signal limit ($|\varphi_N| \ll 1$), the observable signal is proportional to the phase, $\delta \langle I_z \rangle \approx I \varphi_N$, allowing direct readout of the axion interaction.
In the dispersive regime the hyperfine interaction produces a longitudinal shift of the electron splitting proportional to $I_z$, enabling nuclear-to-electron transduction.

The effective Hamiltonian $H_{\text{eff}} = (\omega_e+A I_z) S_z + \omega_N I_z$ indicates that the nuclear polarization acts as a local field modulating the electron Larmor frequency.
Consequently, the axion-driven nuclear dynamics are dispersively upconverted into an electron frequency modulation (FM): $\delta \omega_e(t) = A_{\text{eff}} I |\gamma_N Y_N(\omega_a) B_{a,z}| \cos(\omega_a t+\phi)$.
Here, the signal amplitude is proportional to the nuclear filter gain $|Y_N|$, and we introduce $A_{\text{eff}}=A G_{\text{drv}}$ to generalize the static hyperfine coupling to driven schemes like Hartmann-Hahn resonance or Floquet dressing, where the effective interaction strength is dynamically enhanced.

We express the transduced signal as an effective electron field $B_{\mathrm{eff}}^{(e)} \equiv \delta\omega_{e,0}/\gamma_e$ and define the \rev{intrinsic single-unit} hybrid gain:
\begin{equation}\label{eq:Ghyb}
G_{\text{hyb}}\rev{(\omega_a;\rho_D)}
= \frac{AI\gamma_N}{\gamma_e}|Y_N(\omega_a)|
G_{\text{drv}}\left[1+\mathcal{O}\left(\frac{A}{2\Delta}\right)\right] ,
\end{equation}
where $G_{\text{drv}}=1$ without dressing or driving.
\rev{At donor density $\rho_D$, density enters only through
$\tau_N(\rho_D)$ in $Y_N$; collective and resonator gains are
excluded from Eq.~\eqref{eq:Ghyb}.}
In an electron Ramsey or echo probe with sensitivity function $y_e(t)$ and sinusoidal frequency modulation $\delta\omega_e(t)=\delta\omega_{e,0}\cos(\omega_a t)$, continuous readout yields Bessel sidebands at $\omega_e \pm n\omega_a$ with modulation index $\beta = \delta\omega_{e,0}/\omega_a$.
The axion wind direction $\hat{\bm{u}}_a$ is fixed in the Galactic frame while the laboratory quantization axis rotates with Earth, so the projection $R(t)\hat{\bm{u}}_a\cdot\hat{\bm{z}}$ generates a sidereal component at $\Omega_{\star}$ and annual sidebands at $\Omega_{\star} \pm \Omega_{\oplus}$~\cite{Xiangjun2025_2, supp}.
After upconversion, the modulation spectrum retains the characteristic sidereal triplet and annual sidebands, providing an instrument-independent astrophysical signature.

We quantify the resulting signal-to-noise performance in the electron and nuclear channels.
\rev{We distinguish the coherent duration of an individual sensing block,
bounded by $T_2^{(e,N)}$ and $\tau_a$, from the accumulated effective
integration time $T_{\rm int}^{(e,N),\rm eff}$, which also includes the duty
cycle and the adopted rule for combining repeated coherent segments.}

In the Ramsey limit $\omega_a \tau_N \ll 1$, the nuclear response is approximately flat over the interrogation time, while echo/CPMG sequences produce narrowband peaks.
Long nuclear coherence, strong hyperfine coupling, and low electron readout noise allow the upconversion scheme to surpass direct nuclear readout whenever $T_{\text{eff}}^{(e)}$ and $\tau_a$ are large enough to support coherent integration.
Including both readout noise floors, \rev{the intrinsic readout-channel ratio at fixed donor density is}
\begin{equation}\label{eq:SNR_frac}
\rev{\left.\frac{\text{SNR}_e}{\text{SNR}_N}\right|_{\rho_D}
= G_{\text{hyb}}(\omega_a;\rho_D)
\frac{\eta_B^{(N)}}{\eta_B^{(e)}}
\sqrt{\frac{T_{\rm int}^{(e),\rm eff}}{T_{\rm int}^{(N),\rm eff}}} \,,}
\end{equation}
\rev{Here $\eta^{(N,e)}_B$ are readout-channel magnetic sensitivities;
all factors are evaluated at the same $\rho_D$, before optional
collective and resonator resources~\cite{supp}.  Hybrid readout wins when
the right-hand side exceeds unity.}
\begin{figure}[H]
\centering
\includegraphics[width=0.94\columnwidth]{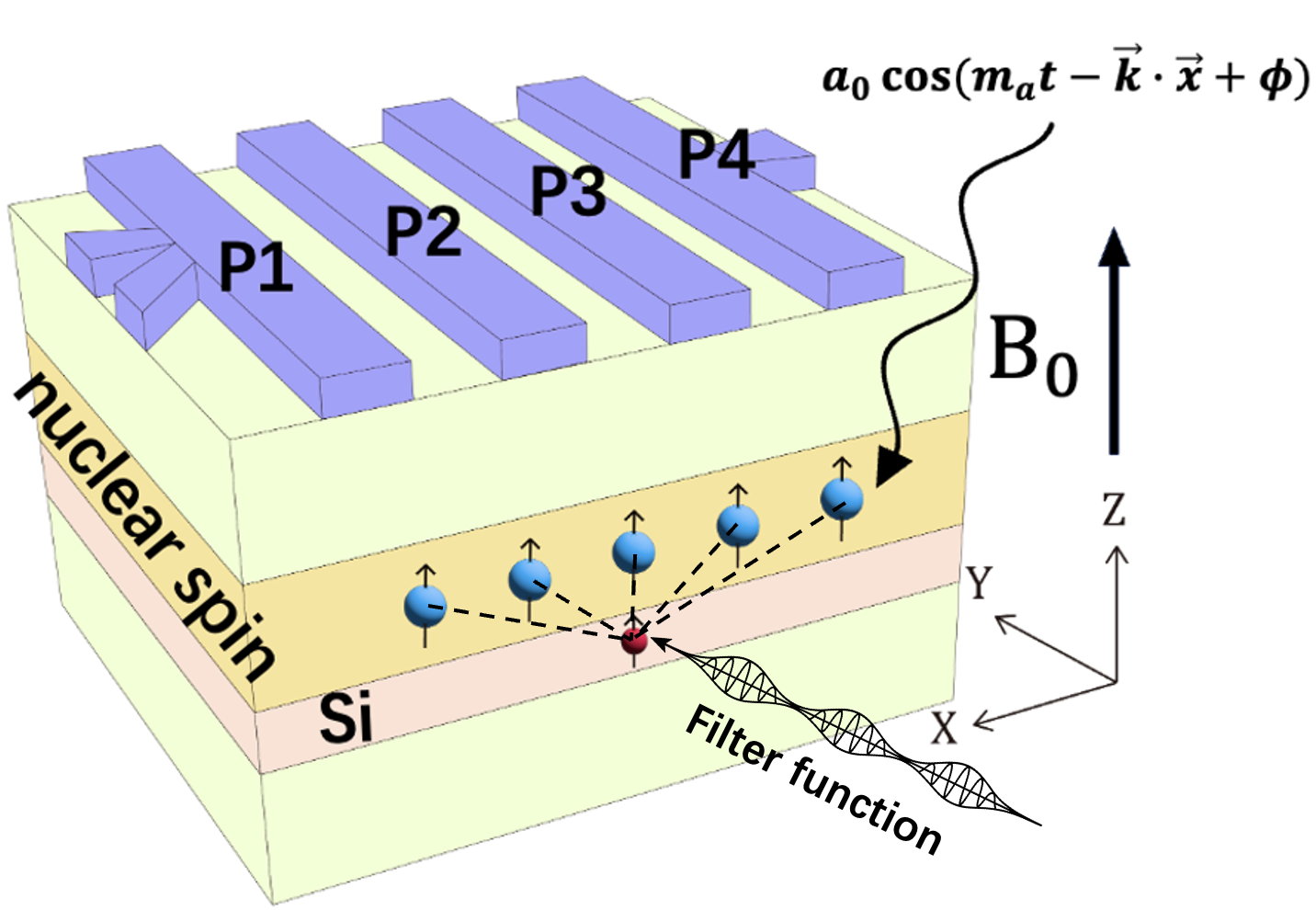}
\caption{
\rev{Single-unit-cell schematic; the sensor uses an array of $N$ coupled units. Local contact hyperfine coupling enables fast, high-bandwidth readout of the axion-sensitive donor nuclear spin (blue) by the gate-defined electron spin (red). The static field $B_0=B_z\hat z$ sets the quantization axis, and gates P1\textendash{}P4 schematically define confinement. Compatible silicon-donor, electron- and hole-spin, and CMOS implementations are described in Refs.~\cite{Pla2014,Muhonen2018,Huang2019,Hendrickx2021,Liles2024,Hendrickx2024,Ik2024,Guangchong2025,Chunhui2026,Gonzalez2021,Veldhorst2014,Veldhorst2015,Maurand2016,Cifuentes2024,Steinacker2025}.}}
\label{fig:hybrid_sensor}
\end{figure}

\textit{Experimental proposal.-}
The device shown in Fig.~\ref{fig:hybrid_sensor} is a chip-integrated hybrid nuclear\textendash{}electronic spin sensor designed for axion and ALP wind searches. All functional elements, including the donor ensemble, electrostatic control gates, microwave drive lines, and dispersive readout circuitry, are integrated on a single cryogenic platform. The chip operates inside radio-frequency and magnetic shielding to suppress ambient electromagnetic interference and low-frequency magnetic-field noise. 
Such noise sources are particularly critical because the axion signal is expected to exhibit sidereal modulation at frequency $\Omega_\star$ and annual sidebands at $\Omega_\star \pm \Omega_\oplus$. 

Suppressing environmental magnetic fluctuations at these frequencies is therefore essential to avoid systematic backgrounds that could mimic the expected modulation pattern. The device is thermally anchored to a dilution refrigerator stage at $T \lesssim 100~\mathrm{mK}$, which ensures long electron and nuclear coherence times, reduces thermal spin polarization noise, and stabilizes dispersive electron readout~\cite{Giardino2022,Kiene2025}.
Superconducting NbTi wiring together with filtered twisted-pair lines is used inside the shielded enclosure to suppress Johnson noise, attenuate high-frequency leakage, and minimize microwave phase noise in the control and readout chain. Under these conditions the dominant sensitivity limitation is set by quantum projection noise and amplifier noise rather than technical electromagnetic 
backgrounds.

Si:$^{209}$Bi gives the largest ideal hyperfine-mediated gain because $A/2\pi=1.475$~GHz, whereas a proof-of-principle experiment could instead use $^{28}$Si:$^{31}$P. The smaller hyperfine interaction of $^{28}$Si:$^{31}$P ($A/2\pi=117$~MHz) reduces the hybrid gain in Eq.~\ref{eq:Ghyb}, but the platform offers mature single-donor control, high-fidelity spin-to-charge readout, and well-established coherence benchmarks. Existing $^{28}$Si:$^{31}$P platforms could therefore validate the hyperfine-mediated transduction and modulation fingerprint before scaling to high-gain Bi ensembles. A comparison of these materials is given in the Supplemental Material~\cite{supp}.

\textit{Results.-}
Fig.~\ref{fig:axion_5sigma_compare_ent_res} presents a sensitivity projection that includes optional collective enhancement and resonator-assisted gain.
\rev{The projection uses
$G_{\rm tot}^{(e)}\equiv\mathcal G_{\rm geom}F_{\rm enh}[N_{\rm eff}(\rho_D)]
G_{\rm res}(Q)G_{\rm hyb}(\omega_a;\rho_D)$, so density enters
separately through $N_{\rm eff}$ and $Y_N$.  Resource and geometry assumptions
are listed in the Supplemental Material~\cite{supp}.}
Single-donor or small-array experiments can validate the hyperfine-mediated upconversion and modulation fingerprint without collective or resonator enhancement.

\begin{figure}[H]
\centering
\includegraphics[width=0.98\columnwidth]{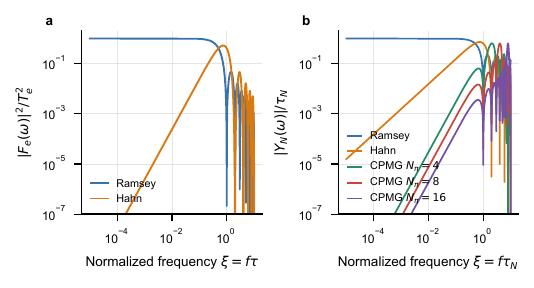}
\caption{
\rev{(a) Normalized Ramsey and Hahn electron filters, shown for reference; the pulsed readout uses $F_e(0)$. (b) Nuclear toggling-filter amplitude for $N_\pi=4,8,16$, with CPMG passbands near $\xi=f\tau_N\simeq N_\pi/2$.}}
\label{fig:filter}
\end{figure}

\begin{figure}[H]
\centering
\includegraphics[width=0.86\columnwidth]{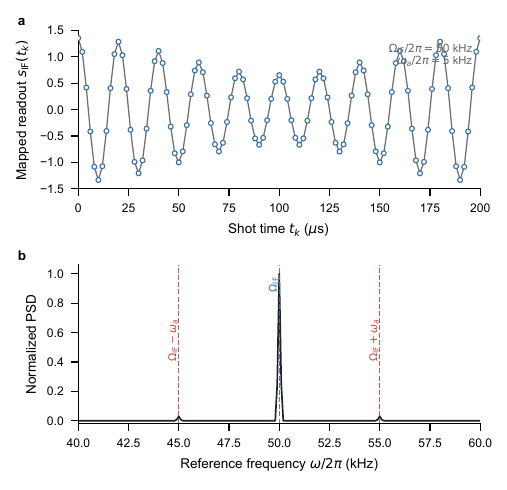}
\caption{
\rev{(a) Shot-resolved electron samples after repeated nuclear sensing and phase mapping, displayed about $\Omega_{\rm IF}/2\pi=50$~kHz with modulation $\omega_a/2\pi=5$~kHz. (b) Spectrum of the same record. This pulsed heterodyne visualization assumes neither spin locking nor a second electron filter at $\omega_a$.}}
\label{fig:spinlock_axion}
\end{figure}

\begin{figure}[H]
\centering
\includegraphics[width=0.96\columnwidth]{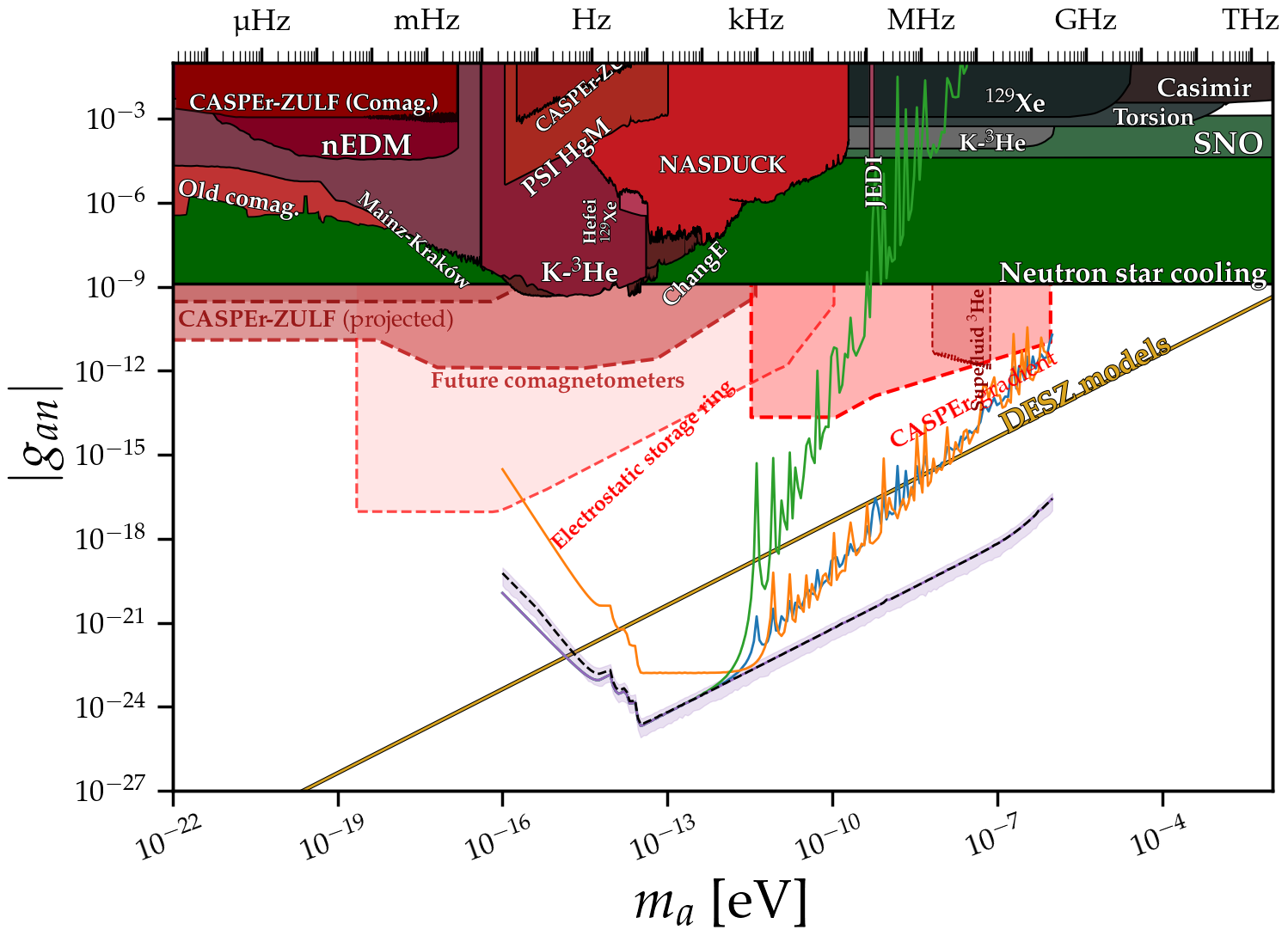}
\caption{
\rev{Projected local $5\sigma$ sensitivity to $g_{an}\equiv m_N g_{aNN}$ for ${}^{209}$Bi under Ramsey, Hahn echo, XY8, and a separate continuous-drive benchmark. This is an idealized per-mass-point benchmark, not an exclusion. Purple shading shows simulated noise spread. The Supplemental Material lists device/noise assumptions; DFSZ and existing limits, partly from Ref.~\cite{AxionLimits}, are shown for reference.}}
\label{fig:axion_5sigma_compare_ent_res}
\end{figure}

To estimate the axion-search sensitivity of the proposed sensors, we normalize the hybrid gain of each isotope to the sensitivity baseline introduced in Ref.~\cite{Xiangjun2025}.
Building on Eq.~\eqref{eq:Ghyb}, we evaluate the nuclear filter $Y_N(\omega)$ together with the quasi-static electron mapping response $F_e(0)$ for a Ramsey/Hahn/CPMG sequence bank and determine the corresponding scan bands from the nuclear passbands.
\rev{At fixed density and with optional resources held fixed, Eq.~\ref{eq:SNR_frac}
therefore isolates the intrinsic hyperfine advantage of $^{209}$Bi.}

\begin{equation}
H_p(\omega;\tau_N,N_\pi)=R_e Y_N(\omega;\tau_N,N_\pi),
\qquad R_e\equiv F_e(0)/T_e.
\end{equation}
This is the pulsed spectral response, rather than $F_e(\omega)Y_N(\omega)$. The electron filter in Fig.~\ref{fig:filter}(a) is therefore shown only for reference. For the standard CPMG convention, the nuclear passband is centered near $f_c\simeq N_\pi/(2\tau_N)$ with a width of order $1/\tau_N$. The nuclear filter $Y_N$ defines the tunable detection band, while $F_e(0)$ sets the electron-readout contrast. The corresponding shot-resolved pulsed-heterodyne readout and its spectrum are shown in Fig.~\ref{fig:spinlock_axion}.

\rev{Within this bounded window, the projected pulsed matched-filter
signal-to-noise ratio is
\begin{equation}
{\rm SNR}^{\rm tot}_e(\omega_a,\rho_D)=
\frac{G_{\rm tot}^{(e)}|B_{a,z}|}{\eta_B^{(e)}}
\sqrt{T_{\rm int}^{(e),\rm eff}}\, .
\end{equation}
The intrinsic frequency dependence is carried by
$G_{\rm hyb}\propto|Y_N[\omega_a;\tau_N(\rho_D)]|$;
the electron factor is the quasi-static mapping response $R_e=F_e(0)/T_e$,
not a second filter at $\omega_a$.  The scan constraints and pulse-selection
rules are given in the Supplemental Material~\cite{supp}.}

\rev{Supplemental Material Fig.~1 shows the corresponding $A$- and
$\tau_N$-dependence of the intrinsic transduction.}

\textit{Conclusion.-}
We propose a hybrid nuclear\textendash{}electronic upconversion scheme for axion-wind sensing while retaining sidereal and annual directional signatures. Using a long-lived, axion-coupled nuclear ensemble and hyperfine transfer to a fast electronic readout, the method is compatible with silicon-donor spin platforms and standard control and readout methods. Monte Carlo simulations indicate the largest gain for Si:$^{209}$Bi donors. With collective enhancement and resonator coupling, the idealized benchmark overlaps part of the DFSZ reference region over $m_a=10^{-16}$\textendash{}$10^{-10}$~eV.

\textit{Data availability.-}
The data are not publicly available. The data are available from the authors upon reasonable request.

\clearpage
\bibliography{refs}
\end{document}


\begin{CJK*}{UTF8}{}
\title{Supplemental Material: Axion signal searching using hybrid nuclear-electronic spin systems}
\author{Xiangjun Tan ({\CJKfamily{gbsn}\mbox{谈相君}})}
\email{xiangjun.tan.25@ucl.ac.uk}
\affiliation{Department of Physics and Astronomy, University College London, WC1E 6BT London, United Kingdom}
\author{Zhanning Wang}
\affiliation{\rev{Instituto de Ciencia de Materiales de Madrid, Consejo Superior de Investigaciones Cient\'{\i}ficas, Sor Juana In\'es de la Cruz 3, 28049 Madrid, Spain}}
\date{September 22, 2026}

\begin{abstract}
This Supplemental Material provides simulation details that support the results presented in the main text.
\end{abstract}

\maketitle
\end{CJK*}

\section{Axion\textendash{}nucleus coupling: relativistic form, NR reduction, and pseudo-field}
\label{subsec:axion_nucleus_coupling}

{\color{black}
We use the dimensionful derivative axion\textendash{}nucleon coupling
$g_{aNN}\equiv C_N/f_a$, with $[g_{aNN}]={\rm energy}^{-1}$, and define
the dimensionless coupling plotted in Fig.~4 of the main text as
$g_{an}\equiv m_N g_{aNN}$.  The interaction is
\begin{align}
\mathcal{L}_{\rm int}
&=\ \frac{g_{aNN}}{2}\,\partial_\mu a\;\bar N \gamma^\mu\gamma^5 N ,
\label{eq:L_derivative}
\end{align}
where $N$ is the nucleon field and $g_{aNN}$ collects model-dependent low-energy constants (e.g.\ $g_{aNN}=C_N/f_a$ in DFSZ scenarios). In the nucleon
nonrelativistic (NR) limit the axial current satisfies
\begin{align}
\bar N \gamma^i\gamma^5 N &\;\to\; 2\,S_N^i + \mathcal{O}(|\bm{p}|/m_N), 
\qquad
\bar N \gamma^0\gamma^5 N \;\to\; \mathcal{O}(|\bm{p}|/m_N),
\end{align}
so that spatial gradients dominate over time gradients. The interaction Hamiltonian
density becomes
\begin{align}
\mathcal{H}_{\rm int}
&=\ -\,\mathcal{L}_{\rm int}
\ \to\ -\,\frac{g_{aNN}}{2}\,(\nabla a)\cdot \bar N \boldsymbol{\gamma}\gamma^5 N
\ \simeq\ -\,g_{aNN}\,(\nabla a)\cdot\bm{S}_N
=-\,\frac{g_{an}}{m_N}\,(\nabla a)\cdot\bm{S}_N ,
\label{eq:H_NR_density}
\end{align}
where we used $\bm{S}_N$ as the spin operator and kept the leading NR term
(the subleading piece $\propto \partial_t a$ is velocity suppressed).
}

The discussion above applies to both the neutron and proton cases, with the corresponding modifications.
The total Hamiltonian is
\begin{equation}
H_{\text{aN}} = -\frac{C_A}{f_a} \nabla a \cdot \bm{I} - \frac{C_A}{f_a} \frac{\partial_t a}{m_N} \bm{p}_N\cdot\bm{I} \,,
\end{equation}
where $\bm{p}_N$ is the nuclear momentum operator, and the effective coupling is
\begin{equation}
C_A=\frac{C_p\left\langle\boldsymbol{S}_p \cdot \boldsymbol{I}\right\rangle+C_n\left\langle\boldsymbol{S}_n \cdot \boldsymbol{I}\right\rangle}{I}\,,
\label{eq:cc}
\end{equation}
where $C_p$ and $C_n$ are the proton and neutron coupling constants, respectively, and $\boldsymbol{S}_p$ and $\boldsymbol{S}_n$ are the corresponding total spin operators inside the nucleus.
Because the nucleus is confined in the solid, its momentum $\bm{p}_N$ is very small, so the second term can be safely neglected.

\subsection{Effective nucleon spin content for donor nuclei (Bi, Sb, P)}
The axion-nucleon derivative interaction at the nuclear level can be parameterized by an effective
dimensionless coefficient $C_A^{(A)}$ for a given nucleus $A$ (spin $I$), defined by the proton/neutron
spin content in the nuclear ground state as in Eq.~\ref{eq:cc}, where $\langle S_{p,n}\rangle \equiv \langle J,M_J{=}J|S_{p,n,z}|J,M_J{=}J\rangle$ are the total proton/neutron spin expectation values in the stretched state, as conventionally quoted in the spin-dependent nuclear-response literature. With this convention the axion pseudo-field entering Eq.~\eqref{eq:HaN_Beff} is obtained by the replacement $g_{aNN}\rightarrow C_A^{(A)}/f_a$ (or equivalently by using $C_A^{(A)}$ in the mapping from nucleon-level to nucleus-level couplings).

For convenience we also define the normalized spin fractions
$a_p\equiv \langle S_p\rangle/I$ and $a_n\equiv \langle S_n\rangle/I$, so that $C_A^{(A)} = C_p a_p + C_n a_n$.
Table~\ref{tab:spin_content} summarizes representative values for donor-relevant nuclei. The key point
is that Bi and Sb isotopes are predominantly proton-spin nuclei (odd $Z$), while $^{31}$P is also
largely proton-dominated; neutron-spin contributions are sub-leading in these examples.

\begin{table}[htb]
\centering
\caption{Proton and neutron spin expectation values and normalized fractions
for donor-relevant nuclei. Values for $^{209}$Bi and $^{31}$P are taken from
Ref.~\cite{Stadnik2015_EPJC}, and those for the Sb isotopes
from Ref.~\cite{universe2022}.}
\label{tab:spin_content}
\begin{tabular}{lcccccc}
\hline\hline
Nucleus & $I$ & $\langle S_p\rangle$ & $\langle S_n\rangle$ 
& $a_p$ & $a_n$ \\
\hline
$^{209}$Bi & $9/2$ & $-0.232$ & $-0.177$ 
& $-5.16\times10^{-2}$ & $-3.93\times10^{-2}$ \\
$^{121}$Sb & $5/2$ & $0.43887$ & $0.04041$ 
& $0.17555$ & $0.01616$ \\
$^{123}$Sb & $7/2$ & $0.272$ & $0.013$ 
& $0.0777$ & $0.0037$ \\
$^{31}$P & $1/2$ & $0.323$ & $0.177$ 
& $0.646$ & $0.354$ \\
\hline\hline
\end{tabular}
\end{table}

For $^{209}$Bi and $^{31}$P we adopt the semi-empirical magnetic-moment-constrained analysis of~\cite{Stadnik2015_EPJC}. For the Sb isotopes we use the large-scale shell-model calculations of~\cite{universe2022}. Representative values are summarized in Table~\ref{tab:spin_content}.
All donor nuclei considered here are odd-$Z$ systems and are therefore
predominantly proton-spin dominated, although neutron contributions
are non-negligible in $^{209}$Bi and $^{31}$P.

{\color{black}
For a spin-$I$ nucleus, define the effective derivative coupling
$g_{aA}\equiv a_p g_{app}+a_n g_{ann}=C_A^{(A)}/f_a$, where
$a_{p,n}=\langle S_{p,n}\rangle/I$.  Writing the total spin operator as
$\bm I$, the leading nuclear Hamiltonian is
\begin{align}
H_{aN}(t)
&=\ -\,g_{aA}\,(\nabla a)\cdot\bm{I}\,,
\label{eq:H_NR_spin_vector}
\end{align}
which is equivalent to a Zeeman coupling in a pseudo-magnetic field.
Dividing by $\hbar$ and introducing the nuclear gyromagnetic ratio
$\gamma_N$, one obtains
\begin{align}
\frac{H_{aN}(t)}{\hbar}
&=\ -\,\gamma_N\,\bm{I}\cdot\bm{B}^{(N)}_a(t), 
\qquad
\bm{B}^{(N)}_a(t)
=\ \frac{g_{aA}}{\hbar\,\gamma_N}\,\nabla a(t,\bm{r}) .
\label{eq:HaN_Beff}
\end{align}
Equation~\eqref{eq:HaN_Beff} identifies the axion wind pseudo-field that drives
the nuclear Zeeman response. For a coherently oscillating field with the SHM kinematics,
\begin{align}
a(\bm{r},t)
&=a_0\cos(\omega_a t-\bm{k}\cdot\bm{r}+\phi_a),
\qquad \omega_a \simeq m_a ,
\\
\nabla a(\bm{r},t) &=\bm{k} a_0\sin(\omega_a t-\bm{k}\cdot\bm{r}+\phi_a), \qquad |\bm{k}| = m_a v_a ,
\end{align}
and the SHM amplitude is fixed by the local density,
\begin{align}
a_0=\frac{\sqrt{2\rho_{\rm DM}}}{m_a}
\quad\Rightarrow\quad
\bigl|\nabla a\bigr|
=\ m_a v_a a_0
=\ v_a\,\sqrt{2\rho_{\rm DM}} \, .
\label{eq:grad_amp}
\end{align}
Therefore the pseudo-field magnitude is
\begin{align}
\bigl|\bm{B}^{(N)}_a\bigr|
&=\ \frac{|g_{aA}|}{\hbar\,\gamma_N}\;
v_a\,\sqrt{2\rho_{\rm DM}}\ ,
\label{eq:Ba_magnitude}
\end{align}
and the component that couples in the secular high-field limit is its projection on the
quantization axis $\hat{\bm{u}}\parallel \bm{B}_0$,
\begin{align}
B_{a,z}(t)
&=\ \hat{\bm{u}}\cdot\bm{B}^{(N)}_a(t)
=\ \bigl|\bm{B}^{(N)}_a\bigr|\;p(t),
\qquad
p(t)\equiv \hat{\bm{u}}\cdot\hat{\bm{k}}(t),
\label{eq:Ba_parallel}
\end{align}
where the slow sidereal/annual modulation of $p(t)$ is entirely geometrical.
}

This can be summarized as an effective longitudinal correction (dispersive cross-Kerr shift) to the electron splitting:
\begin{equation}
\delta\omega_e^{(2)}(t) = \frac{A^2[I(I+1)-m_I(m_I+1)]}{4I\Delta} \delta\expval{I_z(t)} \,.
\end{equation}
This is an $\mathcal{O}(1)$ matrix-element factor that depends on the populated nuclear state.

\subsection{Secular-approximation error for \texorpdfstring{$^{209}$Bi}{Bi-209}: a Schrieffer-Wolff estimate}
\label{subsec:secular_error_bi}

We assess the accuracy of the secular mapping used in the main text for
$^{209}$Bi (Si:Bi, $S=1/2$, $I=9/2$). 
Write the electron-nucleus Hamiltonian as
\begin{equation}
H = \omega_e S_z + \omega_N I_z + A(S_z I_z + \tfrac12 S_+ I_- + \tfrac12 S_- I_+) + H_{aN}(t),
\end{equation}
and denote the Zeeman detuning by $\Delta\equiv\omega_e-\omega_N$ (angular frequencies).
In the high-field secular regime $|\Delta|\gg A$, eliminate the flip-flop term
$V=\tfrac{A}{2}(S_+I_-+S_-I_+)$ by a Schrieffer-Wolff (SW) transformation with
generator $\Omega=\tfrac{A}{2\Delta}(S_+I_- - S_-I_+)$.
Keeping terms up to $\mathcal O(A^2/\Delta)$ gives the effective Hamiltonian
\begin{equation}
H_{\text{eff}}\simeq (\omega_e + A I_z)S_z + \omega_N I_z + \frac{A^2}{2\Delta} (S_z - I_z),
\label{eq:Heff_SW}
\end{equation}
which augments the secular Ising term $A S_z I_z$ by a small, $A^2/\Delta$-scale
renormalization.  The axion-driven nuclear polarization $\delta \langle I_z(t)\rangle$ then
induces an instantaneous electron frequency shift
\begin{equation}
\omega_e^{\text{inst}}(t) = \omega_e + \left[A + \frac{A^2}{2\Delta}\right] \langle I_z(t)\rangle + \cdots ,
\end{equation}
so that the fractional correction to the secular FM amplitude is
\begin{equation}
\varepsilon_{\text{sec}}\equiv\frac{\delta\omega_e^{(2)}}{\delta\omega_e^{(1)}}
\simeq \frac{A}{2\Delta} .
\label{eq:frac_error}
\end{equation}
Equation~\eqref{eq:frac_error} directly translates to a multiplicative correction of the hybrid gain in the main text: $G_{\text{hyb}}\to G_{\text{hyb}} [1+\varepsilon_{\text{sec}}+\mathcal O((A/\Delta)^2)]$.

For $^{209}$Bi we use $A/h=1.47505~\text{GHz}$ and
$(\gamma_e/2\pi)\simeq 28.0~\text{GHz/T}$,
$(\gamma_n/2\pi)=6.963~\text{MHz/T}$; thus
\begin{align}
\frac{\Delta}{2\pi}
&\simeq \left(\frac{\gamma_e}{2\pi}-\frac{\gamma_N}{2\pi}\right)B
\approx (28.0~\text{GHz/T}) B, \\[4pt]
\Rightarrow\quad
\varepsilon_{\text{sec}}
&\approx \frac{1.475~\text{GHz}}{2\times 28.0~\text{GHz/T}\cdot B}
\simeq \frac{0.0263}{B[\text{T}]} .
\label{eq:num_rule}
\end{align}

\begin{table}[htb]
\centering
\caption{Fractional secular error for Si:Bi using Eq.~\eqref{eq:num_rule}.}
\begin{tabular}{c c c c c c c}
\hline
$B$ (T)              & 0.10 & 0.20 & 0.35 & 0.60 & 1.00 & 1.50 \\
\hline
$\varepsilon_{\text{sec}}$ & 0.26 & 0.13 & 0.075 & 0.044 & 0.026 & 0.018 \\
\hline
\end{tabular}
\label{tab:secular_error_bi}
\end{table}

\subsection{Sidereal and annual modulation of the axion-wind projection}
We work in the standard Sun-centered celestial-equatorial frame (SCCEF) with axes $(\hat{\bm X},\hat{\bm Y},\hat{\bm Z})$, where $\hat{\bm Z}$ is aligned with Earth's rotation axis and $\hat{\bm X}$ points to the vernal equinox. In this inertial frame the (Galactic) axion-wind unit vector is quasi-static on daily timescales and can be written as
\begin{equation}
\hat{\bm u}_a(t)=\hat{\bm u}_0+\varepsilon_\oplus\!\left(\hat{\bm u}_c\cos\Omega_\oplus t+\hat{\bm u}_s\sin\Omega_\oplus t\right),
\qquad \varepsilon_\oplus\sim v_\oplus/v_0,
\label{eq:ua_annual}
\end{equation}
where the $\Omega_\oplus$ term accounts for Earth's orbital velocity modulation.

The laboratory quantization axis $\hat{\bm z}$ is fixed in the lab but rotates in the SCCEF with the local sidereal phase $\theta(t)=\Omega_\star t+\theta_0$. For a fixed lab-axis orientation one may write
\begin{equation}
\hat{\bm z}(t)=\hat{\bm z}_0 + \hat{\bm z}_c\cos\theta(t)+\hat{\bm z}_s\sin\theta(t),
\label{eq:z_sidereal}
\end{equation}
with constant vectors $\hat{\bm z}_{0,c,s}$ determined by the site latitude and the time-independent orientation of the quantization axis in the laboratory.

Combining Eqs.~\eqref{eq:ua_annual} and~\eqref{eq:z_sidereal}, the projection relevant to the axion-induced pseudo-field becomes
\begin{equation}
\hat{\bm u}_a(t)\!\cdot\!\hat{\bm z}(t)
= C_0 + C_\star \cos\!\big(\Omega_\star t+\phi_\star\big)
+ \varepsilon_\oplus\!\left[
C_+\cos\!\big((\Omega_\star+\Omega_\oplus)t+\phi_+\big)
+ C_-\cos\!\big((\Omega_\star-\Omega_\oplus)t+\phi_-\big)
\right] + O(\varepsilon_\oplus^2),
\label{eq:triplet_general}
\end{equation}
where the coefficients $C_0,C_\star,C_\pm$ and phases $\phi_{\star,\pm}$ are geometry-dependent but
time-independent. Equation~\ref{eq:triplet_general} shows that Earth's rotation produces a sidereal
component at $\Omega_\star$, while the annual modulation generates sidebands at
$\Omega_\star\pm\Omega_\oplus$, yielding the characteristic sidereal triplet fingerprint.

\section{Scan Constraints and Filter Optimization}

{\color{black}
The scannable axion frequency range is constrained by the nuclear coherence
time and the maximum achievable $\pi$-pulse rate.  For a standard CPMG
sequence with $N_\pi$ pulses at
$t_j=(j-\tfrac12)\tau_N/N_\pi$, adjacent pulses are separated by
$\tau_N/N_\pi$ and the principal passband is centered at
\begin{equation}
f_c\simeq\frac{N_\pi}{2\tau_N}.
\end{equation}
Finite pulse duration requires $\tau_N/N_\pi\gtrsim t_\pi$, and therefore
$f_c\lesssim1/(2t_\pi)$.  For a chosen
$\tau_N^\star\le T_{2,N}$, the nearest even pulse setting is
\begin{equation}
N_\pi^\star=\operatorname{round}_{\rm even}(2f_a\tau_N^\star).
\end{equation}

For the pulsed protocol, the resolved spectral response is
\begin{equation}
H_p(\omega;\tau_N,N_\pi)
=R_e Y_N(\omega;\tau_N,N_\pi),
\qquad R_e\equiv F_e(0)/T_e .
\end{equation}
Thus the dynamic electron filter $F_e(\omega_a)$ is not applied a second
time after nuclear phase mapping.
}

\section{Derivation of the Hybrid Readout Signal-to-Noise Ratio}

{\color{black}
For a nuclear sensing block beginning at $t_k$, the mapped electron-frequency
shift is
\begin{equation}
\delta\omega_{e,k}
=A_{\rm eff}I\gamma_N
\left|Y_N[\omega_a;\tau_N(\rho_{2{\rm D}})]\right|\,|B_{a,z}|
\cos(\omega_a t_k+\Phi_{\rm off}) .
\label{eq:domega}
\end{equation}
The mapped nuclear polarization is quasi-static during the subsequent
electron readout.  Consequently the electron phase for that shot is
\begin{equation}
\phi_{e,k}=\delta\omega_{e,k}F_e(0),
\end{equation}
rather than $\delta\omega_{e,k}F_e(\omega_a)$.  Defining
\begin{align}
B_{e,\rm eff}
&\equiv\frac{\delta\omega_{e,0}}{\gamma_e}
=G_{\rm hyb}(\omega_a;\rho_{2{\rm D}})B_{a,z},\\
G_{\rm hyb}(\omega_a;\rho_{2{\rm D}})
&=\frac{A_{\rm eff}I}{\gamma_e}
\gamma_N\left|Y_N[\omega_a;\tau_N(\rho_{2{\rm D}})]\right| ,
\end{align}
where $\rho_{2{\rm D}}$ is the donor areal density defined below.  This
$G_{\rm hyb}$ is the intrinsic single-unit transduction gain; it contains
neither collective enhancement nor resonator gain.  The matched-filter SNR
for an electron-channel field
sensitivity $\eta_B^{(e)}$ is
\begin{equation}
{\rm SNR}^{\rm int}_e(\rho_{2{\rm D}})
=\frac{|B_{e,\rm eff}|}{\eta_B^{(e)}}
\sqrt{T_{\rm int}^{(e),\rm eff}} .
\end{equation}
Here $T_{\rm int}^{(e),\rm eff}$ includes the duty cycle and the adopted
coherent or incoherent combination of segments; $T_2^{(e)}$ bounds each
electron interrogation but is not identified with the total acquisition
time.

For direct nuclear readout at the same donor density, the same intrinsic
convention gives
\begin{equation}
\left.\frac{{\rm SNR}_e}{{\rm SNR}_N}\right|_{\rho_{2{\rm D}}}
=G_{\rm hyb}(\omega_a;\rho_{2{\rm D}})
\frac{\eta_B^{(N)}}{\eta_B^{(e)}}
\sqrt{\frac{T_{\rm int}^{(e),\rm eff}}
{T_{\rm int}^{(N),\rm eff}}}\, .
\label{eq:SNRratio}
\end{equation}
All factors are evaluated at the same $\rho_{2{\rm D}}$.  Optional collective,
resonator, and geometry resources used in the projection are introduced
separately below, so they are not counted again in Eq.~(\ref{eq:SNRratio}).
}

\begin{figure}[htb]
\centering
\includegraphics[width=0.7\columnwidth]{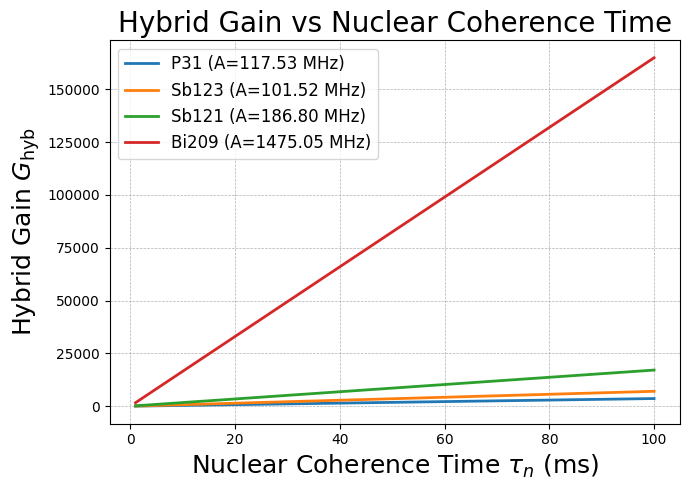}
\caption{
\textbf{Hybrid gain scaling.}
Ideal hybrid gain $G_{\text{hyb}}$ as a function of nuclear coherence time $\tau_N$ for three donor systems.
The scaling is linear in $\tau_N$ and increases with hyperfine coupling strength, with \textsuperscript{209}Bi ($A=1475.05$~MHz) providing more than an order-of-magnitude larger gain than \textsuperscript{31}P and \textsuperscript{121/123}Sb over the plotted coherence-time range.
}
\label{fig:hybrid_gain_vs_tau}
\end{figure}

Figure~\ref{fig:hybrid_gain_vs_tau} compares the intrinsic transduction gain for three donor systems. Over the plotted coherence-time range, $G_{\mathrm{hyb}}$ is linear in $\tau_N$, and the ${}^{209}\mathrm{Bi}$ gain exceeds the P and Sb gains by more than an order of magnitude.

\section{Simulation parameters}
Table~\ref{tab:sim_params} lists the parameters used to generate Fig.~4 of the main text.

\begin{table}[htb]
\caption{
\rev{Simulation parameters used for the best-case $5\sigma$ sensitivity
estimate.  Values correspond to three orthogonal sensing axes, with
$N=10^6$ nuclei per axis, $N_{\rm tot}=3N$, the combined factor
$\mathcal G_{\rm 3ax}=5.4$, and a tunable high-$Q$ resonator.}}
\label{tab:sim_params}
\begin{ruledtabular}
\begin{tabular}{l l l}
\textbf{Quantity} & \textbf{Symbol/Setting} & \textbf{Value} \\
\hline
Donor isotope &  -  & $^{209}$Bi \\
Nuclear spin & $I$ & $9/2$ \\
Hyperfine coupling & $A/2\pi$ & 1475.05~MHz\\
Nuclear gyromagnetic ratio & $\gamma_{N}/2\pi$ &6.963~MHz/T \\
Static magnetic field & $B_0$ & 0.5~T \\
Nuclear drive amplitude & $B_1^{(N)}$ & 10~mT \\
Nuclear decoherence time & $T^{N}_{\mathrm{eff}}$ & 1~s \\
Electron decoherence time & $T^{e}_{\mathrm{eff}}$ & 10~ms \\
Max.~capture window & $T_{\mathrm{cap}}$ & 0.5~s \\
Coherence fraction & $\varepsilon$ & 0.8 \\
Total dwell per mass point & $T_{\mathrm{dwell}}$ & $1~\mathrm{yr}$ \\
Stacking scheme &  -  & $\propto\sqrt{N_{\mathrm{seg}}}$ \\
Magnetic sensitivity & $\eta_B^{(e)}$ & $10^{-10}$~T$/\sqrt{\mathrm{Hz}}$ \\
Magnetic sensitivity & $\eta_B^{(N)}$ & $10^{-8}$~T$/\sqrt{\mathrm{Hz}}$ \\
{\color{black}Three-axis combined factor} & {\color{black}$\mathcal G_{\rm 3ax}$} & {\color{black}$5.4$} \\
Resonator quality factor & $Q$ & $10^{5}$\\
{\color{black}Nuclei per axis} & {\color{black}$N$} & {\color{black}$10^{6}$} \\
{\color{black}Total collective resource} & {\color{black}$N_{\rm tot}=3N$} & {\color{black}$3\times10^{6}$} \\
{\color{black}Entanglement model} & {\color{black}$F_{\mathrm{ent}}(N_{\rm tot})$} & {\color{black}ideal} \\
Dark-matter density & $\rho_{\mathrm{DM}}$ & 0.4~GeV\,cm$^{-3}$ \\
Axion halo velocity & $v_0$ & $10^{-3}c$ \\
\end{tabular}
\end{ruledtabular}
\end{table}

\begin{table}[htb]
\caption{Noise parameters used in the spectral and axion-search sensitivity simulation.}
\label{tab:noise_params}
\begin{ruledtabular}
\begin{tabular}{lccccc}
Noise type &
Amplitude &
$f_{\min}$ (Hz) &
$p_{\rm spike}$ &
$f_{\rm mod}$ (Hz) &
$T$ (K) \\[3pt]
\colrule
White (amplifier floor)  & $1\times10^{-3}$ & \NA  & \NA            & \NA  & \NA \\
Pink ($1/f$ drift)       & $2\times10^{-2}$ & $1$ & \NA            & \NA  & \NA \\
LO/clock phase jitter    & $5\times10^{-4}$ & \NA  & \NA            & \NA  & \NA \\
Control leakage (tones)  & $1\times10^{-4}$ & \NA  & \NA            & 50  & \NA \\
Random telegraph (RTN)   & $5\times10^{-4}$ & \NA  & \NA            & \NA  & \NA \\
Spike/saturation events  & $2\times10^{-3}$ & \NA  & $5\times10^{-3}$ & \NA  & \NA \\
Thermal (base stage)     & \NA               & \NA  & \NA            & \NA  & 10~mK \\
\end{tabular}
\end{ruledtabular}
\end{table}

\section{Density optimization in \texorpdfstring{$\delta$-doped Si:$^{209}$Bi}{delta-doped Si:Bi-209}}

\subsection{Density conventions and ensemble size}
We consider a $\delta$-doped donor layer of active area $\mathcal{A}$ and effective thickness $d_{\rm eff}$.
The donor density may be specified either by an areal density $\rho_{2{\rm D}}$ (cm$^{-2}$) or by an effective
volumetric density $\rho$ (cm$^{-3}$), related by
\begin{equation}
\rho=\frac{\rho_{2{\rm D}}}{d_{\rm eff}},
\qquad
V=\mathcal{A}d_{\rm eff},
\qquad
N=\rho V=\rho_{2{\rm D}}\mathcal{A},
\label{eq:density_def_SN}
\end{equation}
where $V$ is the active sensing volume and $N$ is the number of participating $^{209}$Bi nuclear spins.
Equation~(\ref{eq:density_def_SN}) assumes approximately uniform participation across the layer; in devices with strongly nonuniform coupling, $N_{\mathrm{eff}}$ should be interpreted as an effective ensemble size obtained from an
appropriate spatial weighting. Let the hyperfine coupling between the readout electron and the $j$th Bi nucleus be $A_j$ (set by the electron wavefunction overlap).
The axion-induced response enters through the weighted collective operator
$\tilde I_z \propto \sum_j A_j I_{z,j}$, so that the ensemble enhancement is controlled by an \emph{effective} number of nuclei

\begin{equation}
N_{\mathrm{eff}}
\;\equiv\;
\frac{\left(\sum_j A_j\right)^2}{\sum_j A_j^2},
\label{eq:Neff_def}
\end{equation}%

\subsection{\rev{Minimal density-dependent per-block SNR model}}

In the hybrid architecture, the axion-induced effective field couples to the nuclear spins, while measurement is
performed through the electron channel via hyperfine-mediated transduction.
\rev{We collect the distinct density, resonator, and intrinsic-transduction
factors into the total electron-channel signal-amplitude gain}
\begin{align}
\rev{G_{\rm tot}^{(e)}(\omega_a,\rho_{2{\rm D}},Q)}
&\rev{\equiv
\mathcal G_{\rm geom}\,
F_{\rm enh}[N_{\rm eff}(\rho_{2{\rm D}})]\,
G_{\rm res}(Q)\,
G_{\rm hyb}(\omega_a;\rho_{2{\rm D}}),}
\label{eq:Gtot_SN}\\
\rev{B_{e,{\rm eff}}(\rho_{2{\rm D}})}
&\rev{=
G_{\rm tot}^{(e)}(\omega_a,\rho_{2{\rm D}},Q)\,
B_a^{(N)}.}
\label{eq:Beff_SN}
\end{align}
\rev{Here $F_{\rm enh}$ is the collective signal-amplitude factor and
$G_{\rm res}(Q)$ denotes resonant enhancement.  Neither is absorbed into the
intrinsic $G_{\rm hyb}$.  The three-axis resource and geometry factors used
for Fig.~4 of the main text are those listed in Table~\ref{tab:sim_params}, with
$\mathcal G_{\rm geom}=\mathcal G_{\rm 3ax}$; the one-axis density scaling
sets $\mathcal G_{\rm geom}=1$.}

\rev{The electron readout noise is characterized at the corresponding
readout reference plane by a noise-equivalent magnetic-field sensitivity
$\eta_B^{(e)}$ (T$/\sqrt{\rm Hz}$), before the gains written explicitly in
Eq.~(\ref{eq:Gtot_SN}).  For one coherent sensing block, averaging over
$T_{\rm coh}$ yields}
\begin{equation}
\delta B_e=\frac{\eta_B^{(e)}}{\sqrt{T_{\rm coh}}}.
\label{eq:eta_SN}
\end{equation}
\rev{The corresponding per-block SNR is}
\begin{equation}
\rev{{\rm SNR}_{\rm block}(\rho_{2{\rm D}})}
=
\frac{B_{e,{\rm eff}}(\rho_{2{\rm D}})}{\delta B_e}
=
\frac{
\rev{G_{\rm tot}^{(e)}(\omega_a,\rho_{2{\rm D}},Q)}\,
B_a^{(N)}
}{
\eta_B^{(e)}
}
\sqrt{T_{\rm coh}(\rho_{2{\rm D}})}.
\label{eq:SNR_basic_SN}
\end{equation}

For a narrowband search the control sequence is tuned such that the nuclear filter is evaluated near its passband
maximum at $\omega=\omega_a$. In that case the transduction gain depends primarily on the protocol duration
$\tau_N$,
\begin{equation}
\rev{G_{\rm hyb}(\omega_a;\rho_{2{\rm D}})}
\propto
\rev{\left|Y_N[\omega_a;\tau_N(\rho_{2{\rm D}})]\right|_{\rm peak}}
\sim
\tau_N(\rho_{2{\rm D}}),
\label{eq:Ghyb_peak_SN}
\end{equation}
up to an $\mathcal{O}(1)$ sequence-dependent factor. This peak-aligned form captures the experimentally relevant
frequency-scanning strategy and avoids artifacts that arise when a drifting passband center is compared to a
fixed $\omega_a$.

\subsection{Density-dependent coherence and coherent integration}

Increasing donor density enhances the signal through the number of nuclei $N$, but can reduce the nuclear coherence time through stronger dipolar interactions and local-field broadening. We model the nuclear coherence time by a rate-addition form
\begin{equation}\label{eq:T2N_general_SN}
\rev{\frac{1}{T_{2,N}(\rho_{2{\text{D}}})}}=\frac{1}{T_{2,N}^{(0)}}+\Gamma_{N}^{\rm dip}(\rho_{2{\rm D}})\,,
\end{equation}
where $T_{2,N}^{(0)}$ is the low-density limit and $\Gamma_{N}^{\rm dip}$ is the density-dependent dipolar
dephasing rate.

A minimal effective-thickness parametrization treats the $\delta$ layer as a slab of thickness $d_{\rm eff}$,
so that
\begin{equation}
\Gamma_{N}^{\rm dip}(\rho_{2{\rm D}}) = k_{\rm dip}\rho = k_{\rm dip}\frac{\rho_{2{\rm D}}}{d_{\rm eff}}\,,
\label{eq:T2_density_3D_SN}
\end{equation}
with $k_{\text{dip}}$ in units of cm$^3$/s.
For an ultrathin layer whose dipolar network is effectively
two-dimensional, it can be more natural to write instead
\begin{equation}\label{eq:T2_density_2D_SN}
\Gamma_{N}^{\mathrm{dip}}(\rho_{2\mathrm{D}}) = k_{2\mathrm{D}}\rho_{2\mathrm{D}}\,,
\end{equation}
with $k_{2\mathrm{D}}$ in units of \rev{cm$^2$/s}; the two forms differ only by geometry-dependent factors.

The nuclear sensing block duration is bounded by nuclear coherence, axion coherence, total observation time, and
control constraints. We therefore take
\begin{equation}
\tau_N(\rho_{2{\rm D}})
=
\min\left\{
T_{2,N}(\rho_{2{\rm D}}),\,
\tau_a,\,
T_{\rm obs},\,
\tau_{\rm hw}
\right\},
\label{eq:tauN_SN}
\end{equation}
where $\tau_a$ is the axion coherence time and $\tau_{\rm hw}$ encodes hardware and pulse-sequence limits.

The coherent averaging time in Eq.~(\ref{eq:SNR_basic_SN}) is additionally limited by electron-channel coherence.
We therefore write
\begin{equation}
T_{\rm coh}(\rho_{2{\rm D}})
=
\min\left\{
T_{2,e}(\rho_{2{\rm D}}),\,
\tau_N(\rho_{2{\rm D}}),\,
\tau_a,\,
T_{\rm obs}
\right\}.
\label{eq:Tcoh_SN}
\end{equation}
In donor-based devices $T_{2,e}$ may greatly exceed $\tau_N$, whereas in quantum-dot sensors $T_{2,e}$ can be the
dominant constraint. If needed, density-dependent degradation of electron coherence may be incorporated through a
phenomenological term $1/T_{2,e}=1/T_{2,e}^{(0)}+\Gamma_{e}(\rho_{2{\rm D}})$.

\subsection{Density scaling and practical optimization}

\rev{Combining Eqs.~(\ref{eq:SNR_basic_SN}) and
(\ref{eq:Ghyb_peak_SN}) gives the peak-aligned per-block scaling}
\begin{equation}
\rev{{\rm SNR}_{\rm block}(\rho_{2{\rm D}})}
\propto
F_{\rm enh}(N_{\mathrm{eff}})\,
\tau_N(\rho_{2{\rm D}})\,
\sqrt{T_{\rm coh}(\rho_{2{\rm D}})}.
\label{eq:SNR_peakaligned_SN}
\end{equation}
In the regime where the electron channel does not further restrict the averaging time and
$T_{\rm coh}\simeq\tau_N$, one obtains
\begin{equation}
\rev{{\rm SNR}_{\rm block}(\rho_{2{\rm D}})}
\propto
F_{\rm enh}(N_{\mathrm{eff}})\,
\tau_N^{3/2}(\rho_{2{\rm D}}).
\label{eq:SNR_tauN_SN}
\end{equation}

\rev{For the conservative SQL scaling
$F_{\rm enh}=\sqrt{N_{\rm eff}}\propto\sqrt{\rho_{2{\rm D}}}$,}
\begin{equation}
{\rm SNR}_{\rm SQL}(\rho_{2{\rm D}})
\propto
\sqrt{\rho_{2{\rm D}}}\,
\tau_N^{3/2}(\rho_{2{\rm D}}).
\label{eq:SNR_SQL_SN}
\end{equation}
At low density, $\tau_N\simeq\tau_{\rm hw}$ and ${\rm SNR}_{\rm SQL}\propto \sqrt{\rho_{2{\rm D}}}$.
At higher density where $\tau_N\simeq T_{2,N}(\rho_{2{\rm D}})$, the density scaling follows from the chosen
dipolar model. For example, using Eq.~(\ref{eq:T2_density_3D_SN}) yields $T_{2,N}\propto \rho_{2{\rm D}}^{-1}$ in
the dipolar-dominated regime and therefore
\begin{equation}
{\rm SNR}_{\rm SQL}
\propto
\sqrt{\rho_{2{\rm D}}}\left(\rho_{2{\rm D}}^{-1}\right)^{3/2}
\propto
\rho_{2{\rm D}}^{-1},
\label{eq:SNR_SQL_highdensity_SN}
\end{equation}
implying a true optimum when transitioning from hardware-limited to dipolar-limited operation.

\rev{Equations~(\ref{eq:SNR_basic_SN})\textendash{}(\ref{eq:SNR_tauN_SN}) describe
one coherent block.  The search projection in Fig.~4 of the main text combines repeated blocks
through $T_{\rm int}^{(e),\rm eff}$ and the stacking prescription in
Table~\ref{tab:sim_params}; $T_{\rm coh}$ is not counted a second time.}

A practical operating window is identified near the crossover where dipolar dephasing becomes comparable to the
intrinsic nuclear coherence limit,
\begin{equation}
\Gamma_{N}^{\rm dip}(\rho_{2{\rm D,opt}})
\sim
\frac{1}{T_{2,N}^{(0)}}.
\label{eq:crossover_condition_SN}
\end{equation}
For the effective-thickness model this gives
\begin{equation}
\rho_{2{\rm D,opt}}
\sim
\frac{d_{\rm eff}}{k_{\rm dip}T_{2,N}^{(0)}},
\label{eq:rho2D_opt_3D_SN}
\end{equation}
while for the 2D model,
\begin{equation}
\rho_{2{\rm D,opt}}
\sim
\frac{1}{k_{2{\rm D}}T_{2,N}^{(0)}}.
\label{eq:rho2D_opt_2D_SN}
\end{equation}
If $\tau_{\rm hw}\ll T_{2,N}(\rho_{2{\rm D}})$ over the accessible density range, the experiment remains
hardware-limited and increasing $\rho_{2{\rm D}}$ is beneficial until electron-channel limitations become
significant.

\subsection{Order-of-magnitude dipolar estimate and physical bounds}

An order-of-magnitude estimate follows from the nuclear dipolar interaction between neighboring $^{209}$Bi
spins. For an effective 3D density $\rho$ with characteristic spacing $r\sim \rho^{-1/3}$, the dipolar energy
scale is
\begin{equation}
E_{\rm dip}
\sim
\frac{\mu_0}{4\pi}\frac{\mu_{\rm Bi}^2}{r^3}
\sim
\frac{\mu_0}{4\pi}\mu_{\rm Bi}^2\rho,
\label{eq:Edip_SN}
\end{equation}
where $\mu_{\rm Bi}=\gamma_N\hbar I$ is the nuclear magnetic moment.
Requiring that dipolar-induced dephasing not dominate a target coherence time $T_{2,N}$ gives
\begin{equation}
\frac{E_{\rm dip}}{\hbar}\,T_{2,N}\lesssim 1,
\label{eq:dipolar_condition_SN}
\end{equation}
consistent with the crossover condition in Eq.~(\ref{eq:crossover_condition_SN}) up to geometric factors and the
distinction between effective 2D and 3D dipolar networks.

$\rho_{2{\rm D}}$ is bounded above by atomic-scale areal densities (typically $\sim 10^{15}$\,cm$^{-2}$).
Optimizations predicting $\rho_{2{\rm D,opt}}\gg 10^{15}$\,cm$^{-2}$ therefore indicate that other constraints
(e.g., $\tau_{\rm hw}$ or $T_{2,e}$) remain limiting throughout the physically accessible density range.

\begin{figure}
\centering
\includegraphics[width=0.8\linewidth]{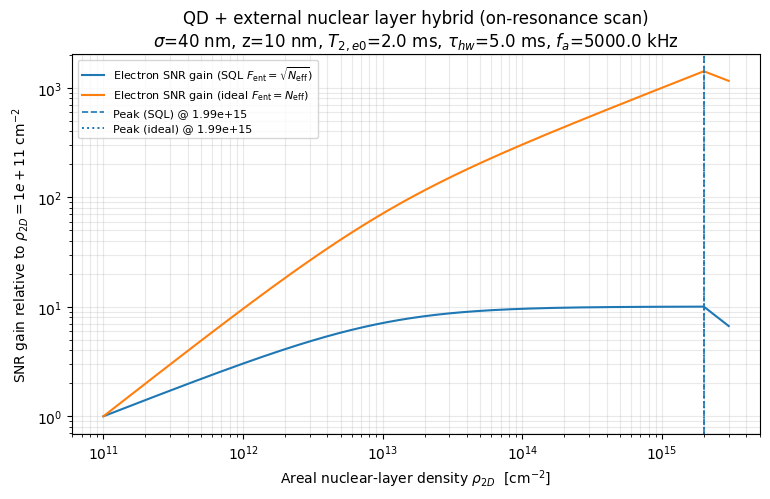}
\caption{
\rev{Density dependence of the hybrid gain for a QD\textendash{}external nuclear-layer architecture under an on-resonance scan.}
The SNR, normalized to its value at \rev{$\rho_{2D}=10^{11}~\mathrm{cm}^{-2}$}, is plotted versus areal nuclear density $\rho_{2D}$. Parameters: \rev{$f_a=5~\mathrm{MHz}$}, $Z=10~\mathrm{nm}$, $T_{2,e}=2.0~\mathrm{ms}$, $\tau_{\mathrm{hw}}=5.0~\mathrm{ms}$ for demonstration. Blue curve: SQL collective scaling $F_{\mathrm{enh}}=\sqrt{N_{\mathrm{eff}}}$. Orange curve: ideal entanglement-enhanced scaling $F_{\mathrm{enh}}=N_{\mathrm{eff}}$. At low density the SNR increases due to collective enhancement, while at high density dipolar-induced nuclear decoherence limits $\tau_N$, leading to saturation and eventual reduction of the SQL-limited response. \rev{The dashed lines mark the simulated response maxima near $\rho_{2D}\simeq2\times10^{15}\,\mathrm{cm}^{-2}$, where $T_{2,N}(\rho_{2D})$ approaches the hardware-limited sensing duration.}
}
\label{fig:density_gain}
\end{figure}
\clearpage
\bibliography{refs}